\documentclass[a4paper,10pt,reqno]{amsart}
\usepackage[utf8]{inputenc}
\usepackage[%
    hyperref,backend=biber, doi=false,url=false,
    sorting=none,style=numeric,backref=true, maxbibnames=7]{biblatex} 
\renewbibmacro{in:}{}

\usepackage{amsmath}
\usepackage{amssymb}
\usepackage{bbm}
\usepackage{bm}
\usepackage{color}
\numberwithin{equation}{section}
\numberwithin{figure}{section}
\usepackage{physics}
\usepackage{graphicx}
\usepackage{epstopdf}
\usepackage{aligned-overset}
\usepackage{hyperref}
\hypersetup{colorlinks=true,linktoc=page,linkcolor=blue,citecolor=red,urlcolor=cyan}

\bibliography{bibliografia}

\usepackage[english]{babel}

\newcommand{\hu}{\hspace{0.6cm}}

\newcommand{\Z}{\mathsf{Z}}
\usepackage{xparse}

\NewDocumentCommand{\dkpara}{O{k} O{\dime}}{\frac{{\rm d} {#1}^{\parallel}}{(2\pi)^{#2}}}
\NewDocumentCommand{\dkd}{O{k} O{\dime}}{ \frac{{\rm d}^{#2} {#1}}{(2\pi)^{#2}}}
\NewDocumentCommand{\dxd}{O{x} O{\dime}}{ {\rm d}^{#2} {#1} }
\NewDocumentCommand{\ad}{O{} O{}}{ \bigl \langle {#1}\bigr\rangle_{{\rm ad}{#2}} }

\makeatletter
\def\@tocline#1#2#3#4#5#6#7{\relax
  \ifnum #1>\c@tocdepth % then omit
  \else
    \par \addpenalty\@secpenalty\addvspace{#2}%
    \begingroup \hyphenpenalty\@M
    \@ifempty{#4}{%
      \@tempdima\csname r@tocindent\number#1\endcsname\relax
    }{%
      \@tempdima#4\relax
    }%
    \parindent\z@ \leftskip#3\relax \advance\leftskip\@tempdima\relax
    \rightskip\@pnumwidth plus4em \parfillskip-\@pnumwidth
    #5\leavevmode\hskip-\@tempdima
      \ifcase #1
       \or\or \hskip 2em \or \hskip 4em \else \hskip 6em \fi%
      #6\nobreak\relax
    \dotfill\hbox to\@pnumwidth{\@tocpagenum{#7}}\par
    \nobreak
    \endgroup
  \fi}
\makeatother
\title{Comment on `Index-free Heat Kernel Coefficients'}
\author{S.~A.~Franchino-Viñas}

\begin{document}

\begin{abstract}
The article by Anton~E.~M.~van de Ven, Class. Quantum Grav. \textbf{15} (1998), is one of the fundamental references for higher-order heat kernel coefficients
in curved backgrounds and with non-abelian gauge connections. 
In this manuscript, we point out two errors and ambiguities in the $\mathsf{a}_5$ coefficient, which may also affect the higher-order ones.
\end{abstract}

\maketitle

%%%%%%%%%%%%%%%%%%%%%%%%%%%%%
%%%%%%%%%%%%%%%%%%%%%%%%%%%%%
%%%%%%%%%%%%%%%%%%%%%%%%%%%%%
%%%%%%%%%%%%%%%%%%%%%%%%%%%%%

\section{Introduction}
Spectral techniques are one of the preferred tools in high energy physics, 
because of their efficiency in the computation of effective actions and vacuum expectation values in nontrivial backgrounds. 
In particular, the heat kernel~\cite{DeWitt:2003} offers a direct, efficient way to analyze the ultraviolet behaviour of a theory, 
carrying this information in its propertime expansion in terms of the Gilkey--Seeley--DeWitt (GSDW) coefficients~\cite{Vassilevich:2003xt}. 
Alternatively, nonperturbative phenomena can be analyzed employing resummations of the heat kernel.

The computation of higher-order GSDW coefficients for a general situation is rather involved and few references tackle this problem,
Ref.~\cite{vandeVen:1997pf} being one of them. 
The importance of having at disposal reliable expressions for these coefficients resides not only in their possible direct application, 
but also in the necessity to have solid cross-checks for analytical developments, such as resummations.

Ref.~\cite{vandeVen:1997pf} presents a method for obtaining the heat kernel coefficients of Laplace-type operators in an index-free notation. 
In this elegant way, its author was able to give the first (implicit) complete expression of the fifth GSDW coefficient for a Laplace operator on a curved manifold,
as well as the sixth-order GSDW coefficient in flat space, including a general (non-)abelian gauge connection in both cases. 

Anyone who has ever attempted this type of computation is aware of how tedious it becomes at such orders; thus, the reader would not be surprised by the presence of some errors in the literature.
In fact, in redoing some of the calculations present in Ref.~\cite{vandeVen:1997pf}, we have found two errors and ambiguities that, while in some sense minor, 
easily turn into headaches for the interested reader.

%%%%%%%%%%%%%%%%%%%%%%%%%%%%%
%%%%%%%%%%%%%%%%%%%%%%%%%%%%%
%%%%%%%%%%%%%%%%%%%%%%%%%%%%%
%%%%%%%%%%%%%%%%%%%%%%%%%%%%%

\section{Discussion}

The first comment concerns the penultimate factor in the expression for $\mathsf{a}_5$, \cite[Eq. (6.1)]{vandeVen:1997pf}, i.e.
 $\frac{4}{21} {\Z}_{(3)}{}^{\dagger} \Z_{(3)}$. 
 After a reading of the article, it becomes clear that the expression is nonsensical, given that, according to the notation of the article, 
 parentheses in the subindices are used to denote a symmetrization followed by a contraction with a product of metric factors, 
 which should be done pair by pair.\footnote{ As an example, ${\Z}_{(2)}:={\Z}_{\mu\nu} g^{\mu\nu}$}.
 One could think that this is just a typographical error, the author meaning ${\Z}_{(3}{}^{\dagger} \Z_{3)}$;
 however, this is not the case. The abovementioned factor should be replaced with
 \begin{align}
  \frac{4}{21} {\Z}_{(3)}{}^{\dagger} \Z_{(3)} \to \frac{4}{21} {\Z}_{(3}{}^{\dagger} \Z_{1)(2)},
 \end{align}
i.e. the indices in the last $\Z$ factor are subject to two different types of symmetrizations and contractions.

The second one, concerns the validity of Eq.~(4.27) of Ref.~\cite{vandeVen:1997pf}, which defines the action of the $\hat{\mathsf{Z}}$ operators 
in flat space, including a nontrivial  gauge connection. 
Note in particular that the term $\hat{\Z}_0\hat{\Z}_0\Z_{1)}$ is involved in the calculation of the $\mathsf{a}_5$ GSDW coefficient.
Applying the second line in \cite[Eq.~(4.27)]{vandeVen:1997pf}, one obtains
\begin{align}
 \text{``}\hat{\Z}_0\hat{\Z}_0\Z_{1)} ={\Z}_0\hat{\Z}_0\Z_{1)}+ 2 \mathsf{Y}_{1}{}^{\nu} \left[ 0 \times \hat{\Z}_{-1}{}^{\nu} \Z_{1)}+ \hat{\Z}_{0)}\Z_{\nu}\right]\text{''}.
\end{align}
An inattentive reader might wonder why we have explicitly written the product of zero times the zeta factors: The problem resides in the fact that $\hat{\Z}_{-1}Z_{1)}$ is not defined in the manuscript.
In fact, \cite[Eq.~(4.27)]{vandeVen:1997pf} gives a definition only for $n\geq0$; in particular, for $n=-1$ one obtains a divergent quantity. 
A direct computation shows that, in this case, the second line in \cite[Eq.~(4.27)]{vandeVen:1997pf} should be intended in the limit $n\to 0$ after replacing every $\hat{\mathsf{Z}}$ factor present in the expression. 
In this process a new contribution is created, which for the term needed in the $\mathsf{a}_5$ calculation reads 
\begin{align}
 \hat{\Z}_0\hat{\Z}_0\Z_{1)} ={\Z}_0\hat{\Z}_0\Z_{1)}+ 4 \mathsf{Y}_{1)}{}^{\nu} \mathsf{Y}_{\nu}{}^{\mu}  \Z_{\mu}+2 \mathsf{Y}_{1)}{}^{\nu}  \hat{\Z}_{0}\Z_{\nu}.
\end{align}
A similar remark applies to the expression that is tacitly assumed in Eq.~(5.42), when a general curved manifold is considered.

Of course, the validity of both these comments can be checked by direct computation of the $\mathsf{a}_5$ GSDW coefficient. 
Using the resummation in Ref.~\cite{Franchino-Vinas:2023}, its explicit expression greatly simplifies, as is shown in App.~\ref{app:coefficients}.
Alternatively, after using integration by parts one can readily compare for example with Ref.~\cite{Fliegner:1997rk} or using the resummed expressions in Ref. \cite{Navarro-Salas:2020oew}.

%%%%%%%%%%%%%%%%%%%%%%%%%%%%%
%%%%%%%%%%%%%%%%%%%%%%%%%%%%%
%%%%%%%%%%%%%%%%%%%%%%%%%%%%%
%%%%%%%%%%%%%%%%%%%%%%%%%%%%%

\section*{Acknowledgments}
The author is indebted to F.~D.~Mazzitelli for several crucial discussions. 
The author acknowledges the support from the Helmholtz-Zen\-trum Dresden-Rossendorf, 
from Consejo Nacional de Investigaciones Cien\-tí\-ficas y Técnicas (CONICET) through the project PIP 11220200101426CO
and from
UNLP through the project 11/X748.

%%%%%%%%%%%%%%%%%%%%%%%%%%%%%
%%%%%%%%%%%%%%%%%%%%%%%%%%%%%
%%%%%%%%%%%%%%%%%%%%%%%%%%%%%
%%%%%%%%%%%%%%%%%%%%%%%%%%%%%

\appendix

\section{First heat kernel coefficients}\label{app:coefficients}
Consider the heat kernel $K(x,x';\tau)$ that satisfies the following equation in a Euclidean, $d$-dimensional space with an abelian gauge connection $A_\mu$ and an arbitrary scalar potential $V$:
\begin{align}\label{eq:HK_eq_EM}
 [\partial_\tau - \nabla^2 +2A_\mu(x) \nabla^\mu + V] K(x,x';\tau)&=0,
 \\
 \label{eq:HK_initial_condition}
K(x,x',0^+)&= \delta(x-x'). 
\end{align}
According to Ref.~\cite{Franchino-Vinas:2023}, its diagonal can be cast as
\begin{align}\label{eq:HK_diagonal}
\begin{split}
K(x,x;\tau)
&=
\frac{e^{-\tau V +\nabla^{\alpha }V \left[ \gamma^{-3} \left({\gamma \tau - 2 \tanh(\tfrac{1}{2} \gamma \tau)}\right)\right]_{\alpha \beta } \nabla^{\beta }V }}
{(4\pi)^{d/2} \;{\det} ^{1/2}\big((\gamma \tau)^{-1} \sinh(\gamma \tau)\big) }  \Omega(x,x;\tau)\Bigg\vert_{\rm coincidence},
\end{split}
\end{align}
where we have defined\footnote{Our notation is $\nabla_{\nu_1\cdots \nu_n}V:= \nabla_{\nu_1}\cdots \nabla_{\nu_n}V$. 
We also use take the indices in second derivatives as matrix indices; 
for a power of a matrix we write $\gamma^n_{\alpha\beta}=\gamma_{\alpha\mu_1}\cdots \gamma_{\mu_n\beta}$.} $\gamma^2_{\mu\nu}:=2\nabla_{\mu\nu}V$. 
In addition, the gauge potential is assumed to be in the Fock--Schwinger gauge, 
so that the scalar potential has both an intrinsic scalar contribution, $\tilde V$, and one resulting from the abelian gauge connection, $V_{EM}$; 
in formulae, we have
\begin{align}\label{eq:VforEM}
 \begin{split}
V:&= \tilde V+V_{EM}(x)
 \\
 V_{EM}(x):&= \nabla_\mu A^\mu-A_\mu A^\mu.
 \end{split}
 \end{align}
The subindex ``coincidence'' in Eq.~\eqref{eq:HK_diagonal} means that every single instance of the gauge field or its derivatives should be replaced by
its coincidence limit (in the Fock--Schwinger gauge~\cite{Franchino-Vinas:2023}), to wit
 \begin{align}\label{eq:gauge_potential}
A^{\nu}{}_{;\mu_1\cdots\mu_n}(x) \to \frac{n}{n+1} F_{\mu_1}{}^{\nu}{}_{;\mu_2\cdots \mu_n}(x).
\end{align}
Note that this is also intended in the prefactor. 

As always, we can perform an expansion of $\Omega$ in the propertime $\tau$, thus defining the modified heat kernel coefficients $c_j$:
\begin{align}\label{eq:HK_Y_expansion}
    \Omega(x,x';\tau)=:\sum_{j=0}^\infty c_j(x,x') \tau^{j-d/2}.
\end{align} 
To enable future comparisons, we write down the explicit expressions for the diagonal of the first five modified heat kernel coefficients; 
all the quantities in the RHS are evaluated at $x$ and Eqs.~\eqref{eq:VforEM} and~\eqref{eq:gauge_potential} are intended:
\begin{align}
c_0(x,x)&=1,\\
c_1(x,x)&=0,\\
c_2(x,x)&=0,\\
c_3(x,x)&= - \tfrac{1}{18} \nabla^{\alpha }V \nabla_{\beta }F_{\alpha }{}^{\beta } -  \tfrac{1}{60} \nabla_{\beta }{}^{\beta }{}_{\alpha }{}^{\alpha }V,
\\
\begin{split}
c_4(x,x)&=
- \tfrac{1}{90} F_{\alpha }{}^{\beta } \nabla^{\alpha }V \nabla_{\gamma }F_{\beta }{}^{\gamma } -  \tfrac{1}{90} \nabla^{\beta \alpha }V \nabla_{\gamma \beta }F_{\alpha }{}^{\gamma }
\\
&\hu\hu+ \tfrac{1}{30} \nabla^{\alpha }V \nabla_{\beta }{}^{\beta }{}_{\alpha }V -  \tfrac{1}{150} \nabla^{\alpha }V \nabla_{\gamma }{}^{\gamma }{}_{\beta }F_{\alpha }{}^{\beta } 
\\
&\hu\hu+ \tfrac{1}{90} \nabla_{\alpha }F^{\alpha \beta } \nabla_{\gamma }{}^{\gamma }{}_{\beta }V -  \tfrac{1}{840} \nabla_{\gamma }{}^{\gamma }{}_{\beta }{}^{\beta }{}_{\alpha }{}^{\alpha }V,
\end{split} 
\\
\begin{split}
 c_5(x,x)&=
 - \tfrac{1}{360} F_{\alpha }{}^{\gamma } F_{\beta \gamma } \nabla^{\alpha }V \nabla^{\beta }V 
 \\
 &\hu\hu -  \tfrac{1}{540} F_{\alpha }{}^{\beta } F_{\beta }{}^{\gamma } \nabla^{\alpha }V \nabla_{\delta }F_{\gamma }{}^{\delta } + \tfrac{1}{90} \nabla^{\alpha }V \nabla_{\gamma }F_{\beta }{}^{\gamma } \nabla^{\beta }{}_{\alpha }V  
 \\
 &\hu\hu -  \tfrac{2}{945} \nabla_{\gamma }F_{\alpha }{}^{\gamma } \nabla_{\delta }F_{\beta }{}^{\delta } \nabla^{\beta \alpha }V -  \tfrac{2}{945} \nabla_{\beta }F_{\alpha }{}^{\gamma } \nabla_{\delta }F_{\gamma }{}^{\delta } \nabla^{\beta \alpha }V  
 \\
 &\hu\hu + \tfrac{1}{120} \nabla^{\alpha }V \nabla^{\beta }V \nabla_{\gamma \beta }F_{\alpha }{}^{\gamma } -  \tfrac{1}{630} \nabla^{\alpha }V \nabla^{\gamma }F_{\alpha }{}^{\beta } \nabla_{\delta \beta }F_{\gamma }{}^{\delta }  
 \\
 &\hu\hu -  \tfrac{1}{630} F^{\alpha \beta } \nabla^{\gamma }{}_{\alpha }V \nabla_{\delta \beta }F_{\gamma }{}^{\delta } + \tfrac{1}{420} \nabla^{\alpha }V \nabla_{\beta }F^{\beta \gamma } \nabla_{\delta \gamma }F_{\alpha }{}^{\delta }  
 \\
 &\hu\hu -  \tfrac{1}{630} \nabla^{\alpha }V \nabla^{\gamma }F_{\alpha }{}^{\beta } \nabla_{\delta \gamma }F_{\beta }{}^{\delta } -  \tfrac{1}{630} F^{\alpha \beta } \nabla^{\gamma }{}_{\alpha }V \nabla_{\delta \gamma }F_{\beta }{}^{\delta }  
 \\
 &\hu\hu + \tfrac{1}{840} \nabla^{\alpha }V \nabla_{\beta }F^{\beta \gamma } \nabla_{\delta }{}^{\delta }F_{\alpha \gamma } + \tfrac{17}{5040} \nabla^{\beta }{}_{\alpha }{}^{\alpha }V \nabla_{\gamma }{}^{\gamma }{}_{\beta }V  
 \\
 &\hu\hu + \tfrac{1}{840} \nabla_{\gamma \beta \alpha }V \nabla^{\gamma \beta \alpha }V -  \tfrac{1}{3500} \nabla^{\gamma \beta \alpha }V \nabla_{\delta \beta \alpha }F_{\gamma }{}^{\delta }  
 \\
 &\hu\hu -  \tfrac{1}{875} \nabla^{\gamma \beta \alpha }V \nabla_{\delta \gamma \beta }F_{\alpha }{}^{\delta } -  \tfrac{1}{3500} \nabla^{\gamma \beta \alpha }V \nabla_{\delta }{}^{\delta }{}_{\beta }F_{\alpha \gamma }  
 \\
 &\hu\hu + \tfrac{2}{945} F^{\alpha \beta } \nabla_{\gamma }F_{\alpha }{}^{\gamma } \nabla_{\delta }{}^{\delta }{}_{\beta }V -  \tfrac{2}{1575} F_{\alpha }{}^{\beta } \nabla^{\alpha }V \nabla_{\delta }{}^{\delta }{}_{\gamma }F_{\beta }{}^{\gamma }  
 \\
 &\hu\hu -  \tfrac{17}{12600} \nabla^{\beta }{}_{\alpha }{}^{\alpha }V \nabla_{\delta }{}^{\delta }{}_{\gamma }F_{\beta }{}^{\gamma } + \tfrac{1}{210} \nabla^{\beta \alpha }V \nabla_{\gamma }{}^{\gamma }{}_{\beta \alpha }V  
 \\
 &\hu\hu -  \tfrac{1}{630} \nabla^{\beta \alpha }V \nabla_{\delta }{}^{\delta }{}_{\gamma \beta }F_{\alpha }{}^{\gamma } + \tfrac{1}{420} \nabla^{\gamma }{}_{\alpha }F^{\alpha \beta } \nabla_{\delta }{}^{\delta }{}_{\gamma \beta }V  
 \\
 &\hu\hu + \tfrac{1}{280} \nabla^{\alpha }V \nabla_{\gamma }{}^{\gamma }{}_{\beta }{}^{\beta }{}_{\alpha }V -  \tfrac{1}{1960} \nabla^{\alpha }V \nabla_{\delta }{}^{\delta }{}_{\gamma }{}^{\gamma }{}_{\beta }F_{\alpha }{}^{\beta }  
 \\
 &\hu\hu + \tfrac{1}{840} \nabla_{\alpha }F^{\alpha \beta } \nabla_{\delta }{}^{\delta }{}_{\gamma }{}^{\gamma }{}_{\beta }V -  \tfrac{1}{15120} \nabla_{\delta }{}^{\delta }{}_{\gamma }{}^{\gamma }{}_{\beta }{}^{\beta }{}_{\alpha }{}^{\alpha }V.
\end{split}
\end{align}

\printbibliography
\end{document}